\documentclass[twocolumn,prc,preprintnumbers,floatfix,superscriptaddress]{revtex4}
\usepackage{graphicx}
\usepackage{amsmath,amssymb}
\usepackage{bm}
\usepackage{braket}
\usepackage{epsf}
\usepackage{color}
\usepackage[breaklinks=true]{hyperref}

\begin{document}

\preprint{NITEP 30}

\title{Effect of the repulsive core in the proton-neutron potential on
deuteron elastic breakup cross sections}

\author{Yuen Sim Neoh}
\email[]{neohys@rcnp.osaka-u.ac.jp}
\affiliation{Research Center for Nuclear Physics, Osaka University, Ibaraki 567-0047, Japan}
\author{Mengjiao Lyu}
\affiliation{Research Center for Nuclear Physics, Osaka University, Ibaraki 567-0047, Japan}
\author{Yoshiki Chazono}
\affiliation{Research Center for Nuclear Physics, Osaka University, Ibaraki 567-0047, Japan}
\author{Kazuyuki Ogata}
\email[Corresponding author: ]{kazuyuki@rcnp.osaka-u.ac.jp}
\affiliation{Research Center for Nuclear Physics, Osaka University, Ibaraki 567-0047, Japan}
\affiliation{Department of Physics, Osaka City University, Osaka 558-8585, Japan}
\affiliation{
Nambu Yoichiro Institute of Theoretical and Experimental Physics (NITEP),
   Osaka City University, Osaka 558-8585, Japan}

\date{\today}
\begin{abstract}
The role of the short-range part (repulsive core) of the proton-neutron ($pn$) potential in deuteron elastic breakup processes is investigated. A simplified one-range Gaussian potential and the Argonne V4' (AV4') central potential are adopted in the continuum-discretized coupled-channels (CDCC) method. The deuteron breakup cross sections calculated with these two potentials are compared. The repulsive core is found not to affect the deuteron breakup cross sections at energies from 40~MeV to 1~GeV. To understand this result, an analysis of the peripherality of the elastic breakup processes concerning the $p$-$n$ relative coordinate is performed. It is found that for the breakup processes populating the $pn$ continua with orbital angular momentum $\ell$ different from 0, the reaction process is peripheral, whereas it is not for the breakup to the $\ell=0$ continua (the s-wave breakup). The result of the peripherality analysis indicates that the whole spatial region of deuteron contributes to the s-wave breakup.
\end{abstract}

\maketitle

\section{Introduction\label{sec:intro}}

The nucleon-nucleon ($NN$) interaction, the fundamental building block of nuclear physics, has intensively been studied by the phase shift analysis \cite{Arn07,Wor16}, meson theory \cite{Mac01}, chiral effective field theory \cite{Epe09,Mac11}, and lattice QCD \cite{Ish07}. It is well known that the $NN$ interaction has a repulsive core at a short distance. It also contains many spin-dependent terms and among them, the tensor part plays a crucial role in the binding mechanism of deuteron. Many efforts have been devoted to revealing roles of these characteristic features of the $NN$ interaction in many-nucleon systems \cite{carlson_quantum_2015,hagen_coupled-cluster_2014,barrett_ab_2013,lee_lattice_2009,roth_nuclear_2010,hiyama_gaussian_2012,myo_tensor-optimized_2015,lyu_tensor-optimized_2018}.  These have been studied also experimentally via the electron- or proton-induced reactions~\cite{Korover:2014dma, Hen:2014nza, Duer:2018sby, Terashima:2018bwq}. In the same direction, breakup reactions of nuclei will be a possible way of probing the role of the short-range repulsion and tensor-induced attraction.

For many years, breakup reactions of weakly-bound nuclei have been studied theoretically and experimentally. These are mainly motivated by the interest in natures of unstable nuclei and strong couplings with continuum states of fragile systems. Deuteron is the lightest weakly-bound nucleus and its breakup processes have been measured since the early 1980s. The continuum-discretized coupled-channels method (CDCC)~\cite{Kam86,AUSTERN1987125,Yahiro01012012} is one of the most successful reaction models for describing the breakup processes of deuteron and unstable nuclei. Its theoretical foundation was given in Refs.~\cite{PhysRevLett.63.2649,PhysRevC.53.314} and later numerically confirmed \cite{Del07,Upa12,Oga16} via comparisons with Faddeev--Alt-Grassberger-Sandhas (FAGS) theory \cite{Fad60,Alt67}. In most cases, a simplified one-range Gaussian potential \cite{Ohm70} is employed for the $pn$ interaction.

In this study, we consider the deuteron breakup as a possible probe for the abovementioned striking features of the $pn$ interaction. In Ref.~\cite{ISERI1991574}, Iseri and collaborators compared CDCC results with the one-range Gaussian and Reid soft core \cite{Rei68} $pn$ interactions for the cross section and polarization observables in deuteron elastic scattering. The difference is appreciable in tensor analyzing powers but not so significant except for a specific combination of polarization transfer coefficients. According to this finding, we focus on the central part of the $pn$ interaction and use the Argonne V4' (AV4')~\cite{PhysRevLett.89.182501} parameterization as a realistic $pn$ interaction. It should be noted that the role of the tensor and other spin-dependent terms in the Argonne V18 (AV18) interaction~\cite{AV18} are effectively included in the AV4' interaction that only has the central part. For a simple notation, however, we regard the appearance of the short-range repulsive core as the characteristic of the AV4' potential in what follows.

We investigate the effect of the short-range repulsive core on the deuteron breakup cross sections at the deuteron energies from 40~MeV to 1~GeV. Pheripherality of the reaction process regarding the $pn$ relative distance, which is crucial to understand whether the observable reflects the inner part of the $pn$ wave function, is also investigated.

The organization of this paper is as follow. In Sec.~\ref{sec:formulation}, we summarize the method of CDCC, the results are discussed in Sec.~\ref{sec:results}, and finally the conclusion is given in Sec.~\ref{sec:conclusion}.

\section{Formalism\label{sec:formulation}}

\begin{figure}[!h]
\centering\includegraphics{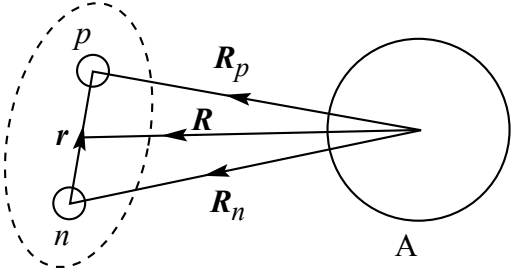}
\caption{Schematics of the three body system in deuteron scattering.}
\label{fig1}
\end{figure}

We describe the deuteron breakup on a target nucleus A by a $p+n+{\rm A}$ three-body model, with assuming A to be inert. The coordinate labels are shown in Fig.~\ref{fig1}.
The three-body Hamiltonian is given by
\begin{equation}
  H=T_{\bm R}+H_{pn}+V_{\rm CL}(R)+V_p(\boldsymbol{R}_p)+V_n(\boldsymbol{R}_n),
  \label{totalH}
\end{equation}
where $T_{\bm X}$ is the kinetic energy operator associated with the coordinate ${\bm X}$, $V_p(\boldsymbol{R}_p)$ and $V_n(\boldsymbol{R}_n)$ are the $p$-A and $n$-A distorting potentials, respectively, and $V_{\rm CL}(R)$ is the Coulomb potential between the center-of-mass (c.m.) of deuteron and A.
The Hamiltonian $H_{pn}$ of the $p$-$n$ system is given by
\begin{equation}
 H_{pn} = T_{\bm r}+V_{pn}(\boldsymbol{r})\label{Hpn},
\end{equation}
where $V_{pn}(\boldsymbol{r})$ is the interaction potential between $p$ and $n$. For the purpose of this investigation, only nuclear breakup is being considered and the intrinsic spin of nucleon is disregarded.

In CDCC, the three-body wave function $\Psi_{JM}$ with the total angular momentum $J$ and its $z$-component $M$ is expanded in terms of the $p$-$n$ eigenstates $\hat{\phi}_{i \ell}$ consisting of the deuteron bound state and discretized continuum states of the $p$-$n$ system:
\begin{align}
  \Psi_{JM}(\boldsymbol{r},\boldsymbol{R})
  &=
  \sum^{i_{\rm max}}_{i=0}\sum^{\ell_{\rm max}}_{\ell=0}\sum^{J+\ell}_{L=|J-\ell|} \hat{\phi}_{i\ell}(r)\hat{\chi}^J_{c}(R)\mathcal{Y}^{JM}_{\ell L}, \label{psi}
  \\
 \mathcal{Y}^{JM}_{\ell L}&=\left[i^\ell Y_\ell(\boldsymbol{\hat{r}})\otimes i^LY_L(\boldsymbol{\hat{R}})\right]_{JM},
\end{align}
where $i$ and $\ell$ are the energy index and the orbital angular momentum of the $p$-$n$ system, respectively; $\hat{\phi}_{00}$ corresponds to the ground state of deuteron. $\hat{\chi}^J_c$ describes the scattering motion of the c.m. of the $p$-$n$ system with respect to A, with $L$ being the relative orbital angular momentum and $c=\{i,\ell,L\}$. The set $\{\hat{\phi}_{i\ell}\}$ satisfies
\begin{equation}
\int d\boldsymbol{r}\,
\hat{\phi}^*_{i'\ell'}(r) Y^*_{\ell' m'}(\boldsymbol{\hat{r}})
 H_{pn} \hat{\phi}_{i\ell}(r) Y_{\ell m}(\boldsymbol{\hat{r}})
 =
\hat{\epsilon}_{i\ell} \delta_{i'i} \delta_{\ell' \ell} \delta_{m'm}
\label{omp}
\end{equation}
and is assumed to form a complete set in a space that is needed for describing a reaction process of interest.

If one inserts Eq.~(\ref{psi}) into the Schr\"odinger equation
\begin{equation}
(H-E)\Psi_{JM}(\boldsymbol{r},\boldsymbol{R})=0
\end{equation}
and multiplies it by $\hat{\phi}^*_{i'\ell'}$ from the left, after the integration over $\boldsymbol{r}$, the following coupled-channels equations for
$\hat{u}^J_c \equiv R \hat{\chi}^J_c$ are obtained:
\begin{align}
  \left(-\frac{\hbar^2}{2\mu_{R}}\nabla^2_R+\frac{\hbar^2}{2\mu_{R}}\frac{L(L+1)}{R^2}+V_{\rm CL}(R)+\hat{\epsilon}_{i\ell}-E\right)\hat{u}^J_c(R) \nonumber \\
=-\sum_{c'}F_{cc'}(R)\hat{u}^J_{c'}(R),
\label{cdcceq}
\end{align}
where $\mu_{R}$ is the reduced mass of the deuteron-A system, $E$ is the total energy, and the form factor $F_{cc'}$ is defined by
\begin{equation}
  F_{cc'}(R) = \left\langle\hat{\phi}_{i\ell}(r)\mathcal{Y}^{JM}_{\ell L}\right|V_p(\boldsymbol{R}_p)+V_n(\boldsymbol{R}_n)\left|\hat{\phi}_{i'\ell'}(r)\mathcal{Y}^{JM}_{\ell' L'}\right\rangle
  \label{F_cc'}.
\end{equation}
Here, the integration is understood to be done for $\bm{r}$ and $\hat{\bm R}$, and use has been made of Eq.~(\ref{omp}).

Equations~(\ref{cdcceq}) are solved under the asymptotic boundary condition of
\begin{equation}
  \hat{u}_c(R)\rightarrow H^{(-)}_{\eta_i,L}(K_iR)\delta_{cc_0}-\sqrt{\frac{K_0}{K_i}}S_{cc_0}H^{(+)}_{\eta_i,L}(K_iR)
  \label{open}
\end{equation}
if $K_i=\sqrt{2\mu_R(E-\hat{\epsilon}_i)}/\hbar$ is real, and
\begin{equation}
  \hat{u}_c(R)\rightarrow -S_{cc_0}W_{-\eta_i,L+1/2}(-2iK_iR)
\end{equation}
if $K_i$ is imaginary. Here, $H^{(+)}_{\eta_i,L}$ ($H^{(-)}_{\eta_i,L}$) is the outgoing (incoming) Coulomb wave function, $W_{-\eta_i,L+1/2}$ is the Whittaker function, and $\eta_i$ is the Sommerfeld parameter. $S_{cc_0}$ in Eq.~(\ref{open}) is the scattering matrix for the transtion to channel $c$ from the incident channel $c_0=\{0,0,J\}$. For more detail, readers are referred to Refs.~\cite{Kam86,AUSTERN1987125,Yahiro01012012}.

\section{Results and discussion\label{sec:results}}

\subsection{\textit{pn} interaction and model space}
\label{subsec:input}

We consider the deuteron scattering on a representative $^{58}$Ni target at incident energies from 40~MeV to 1~GeV. At each energy, the nucleon-target potential is obtained by folding the Melbourne \textit{g}-matrix interaction~\cite{Amo00} with target density similar to the procedures described in Ref.~\cite{PhysRevC.94.044619}. To investigate the role of the short-range repulsive core, we use two $p$-$n$ interactions. One is the AV4' interaction and the other is the one-range Gaussian potential
\begin{equation}
 V_{pn}(r)=-V_0\exp\left(\frac{r^2}{a^2}\right)
\end{equation}
with $V_0=52.10$~MeV and $a=1.812$~fm. The parameters are determined so that the binding energy (2.24 MeV) and the root-mean-square (rms) radius (2.01 fm) of deuteron agree with the values obtained with the AV4' interaction. In what follows, we denote this potential as 1G-av4.
\begin{figure}
\includegraphics[scale=0.9]{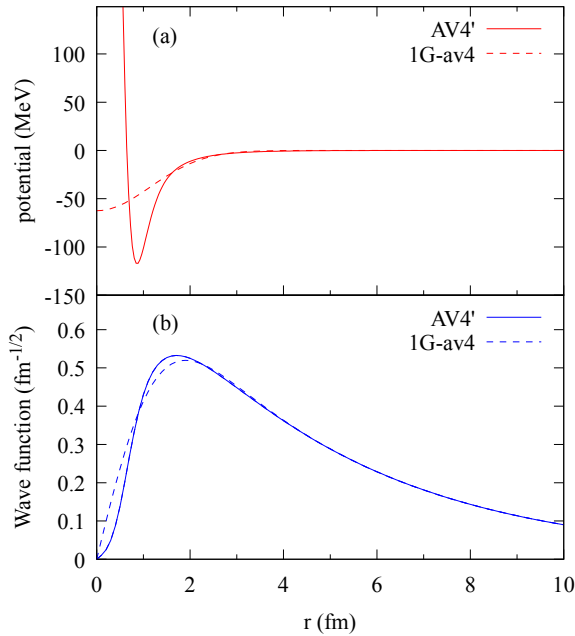}
\caption{
(a) AV4' (solid) and 1G-av4 (dashed) potentials. (b) Radial wave functions of deuteron multiplied by $r$.
}
\label{fig:pot-u_r-2.01}
\end{figure}
In Fig.~\ref{fig:pot-u_r-2.01}(a), we show the AV4' and 1G-av4 interactions by the solid and dashed lines, respectively. The corresponding wave functions of the deuteron ground state multiplied by $r$ are shown in Fig.~\ref{fig:pot-u_r-2.01}(b).

As for the CDCC model space, $p$-$n$ continua with $\ell=0$, 2, and 4 are included, and $r_{\rm max}=R_{\rm max}=60~\text{fm}$. At 80 MeV, the $p$-$n$ states are discretized with momentum bin size $\Delta k$ of 0.05 fm$^{-1}$ up to $k_{\rm max}=1.5~{\rm fm}^{-1}$, and $J_\text{max}=80$; at 1 GeV, $\Delta k=0.25$~fm$^{-1}$ and $k_{\rm max}=3~{\rm fm}^{-1}$, and $J_\text{max}=200$.

\subsection{Breakup cross section}
\label{subsec:breakup}

\begin{figure}[ht]
\includegraphics[scale=1.1]{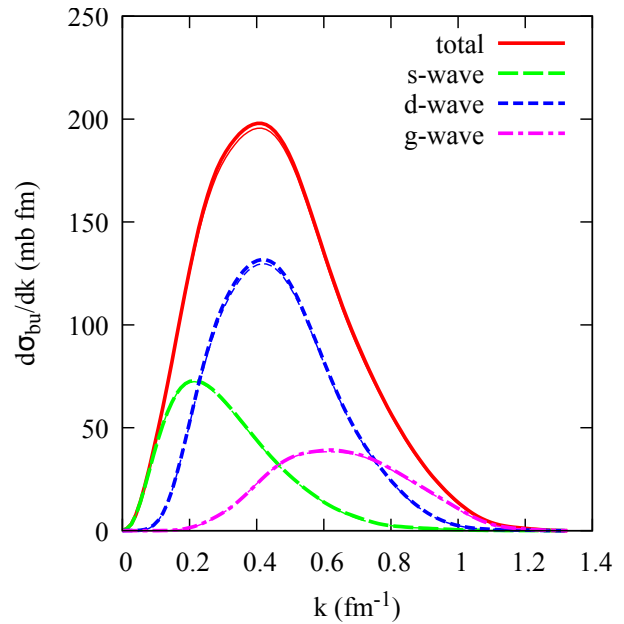}
\caption{\label{fig:80MeV_pbu_micsys}
Differential deuteron breakup cross sections on $^{58}$Ni at 80 MeV as a function of the relative $p$-$n$ momentum (solid lines). The s-, d-, and g-wave components are shown by the dashed, dotted, and dash-dotted lines, respectively. The thick (thin) lines represent the results calculated with the AV4' (1G-av4) potential.}
\end{figure}
\begin{figure}[ht]
\includegraphics[scale=1.1]{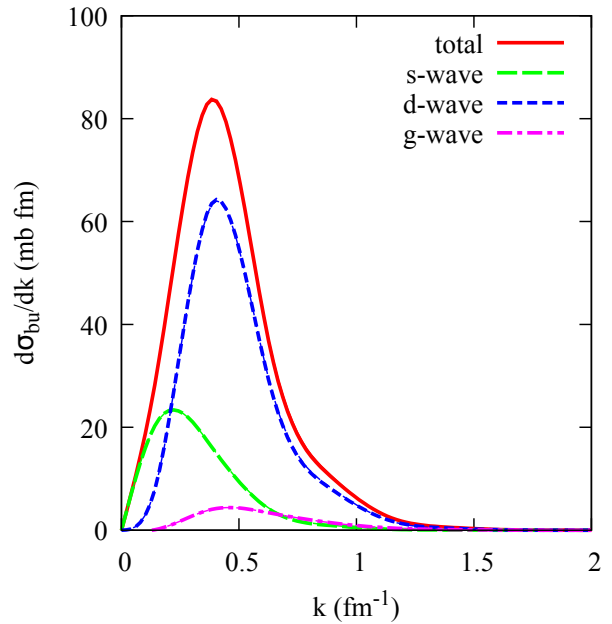}
\caption{Same as Fig.~\ref{fig:80MeV_pbu_micsys} but at 1 GeV.}
\label{fig:1GeV_pbu_micsys}
\end{figure}

Figure~\ref{fig:80MeV_pbu_micsys} shows the differential breakup cross sections at 80~MeV as a function of the $p$-$n$ relative momentum $k$ calculated with the AV4' (thick lines) and 1G-av4 (thin lines) potentials. The s-, d-, and g-wave components are shown by the dashed, dotted, and dash-dotted lines, respectively, and the solid lines are the sum of them. One sees that the difference between the AV4' and 1G-av4 results for each partial-wave component is less than 2\%.
One may expect that at higher incident energies, we will have more chance to
to directly access the short-range part. However, as shown in
Fig.~\ref{fig:1GeV_pbu_micsys}, even at 1 GeV, the AV4' and 1G-av4 potentials do not give an appreciable difference in the breakup cross sections.
The integrated breakup cross sections as well as those breakdown into the partial-wave components are shown in Table~\ref{tab:table2}.
\begin{table*}[ht]
\caption{\label{tab:table2}%
Total breakup cross section and its breakdown into partial-wave components for deuteron on $^{58}$Ni at 80~MeV and 1~GeV.
}
\begin{ruledtabular}
\begin{tabular}{cccccc}
energy &
$V_{pn}$&
\textrm{Total (mb)}&
\textrm{s-wave (mb)}&
\textrm{d-wave (mb)}&
\textrm{g-wave (mb)}\\
\colrule
80 MeV & AV4'   & 106.7 & 27.3 & 58.4 & 21.0\\
       & 1G-av4 & 105.3 & 27.0 & 57.5 & 20.7\\
1 GeV  & AV4'    & 40.8 & 10.5 & 27.9 & 24.3\\
       & 1G-av4  & 40.9 & 10.6 & 27.9 & 24.3
\end{tabular}
\end{ruledtabular}
\end{table*}

Thus, it is found that the short-range repulsive core of the AV4' potential little affects the deuteron breakup cross sections on $^{58}$Ni at 80~MeV and 1~GeV. We have confirmed the same feature at also 40~MeV and 200~MeV (not shown). Furthermore, the negligible difference between the results with the two $V_{pn}$ is found to be robust against the change in the nucleon-A distorting potential.

As seen from Fig.~\ref{fig:pot-u_r-2.01}, the short-range repulsive core in the AV4' potential modifies the inner region of the deuteron wave function. The insensitivity of the breakup cross sections to the difference in $V_{pn}$ may indicate that the deuteron breakup process is peripheral concerning the $p$-$n$ relative distance $r$.

\subsection{Peripherality of deuteron breakup process}
\label{subsec:periph}

To investigate the peripherality of the deuteron breakup, we follow the idea of the asymptotic normalization coefficient (ANC) method~\cite{AMN.NKT,PhysRevLett.73.2027} but with no intention to determine the ANC. In the asymptotic region, i.e., beyond the range $r_N$ of $V_{pn}$, the deuteron wave function becomes
\begin{equation}
  \varphi(r)\xrightarrow{r>r_N} b \exp(-\kappa r),
  \label{I_asymptotic_h}
\end{equation}
where $\kappa=(2\mu_{pn}\varepsilon/\hbar^2)^{1/2}$ with $\mu_{pn}$ being the $p$-$n$ reduced mass and $\varepsilon$ the deuteron binding energy. $b$ is the ANC if a realistic $\varphi$ is used. In the present investigation, however, $b$ is regarded to be just a constant. If the deuteron breakup is peripheral, the breakup cross section $\sigma_{\rm bu}$ is proportional to $b^2$. Then, if we change $V_{pn}$, $b$ and $\sigma_{\rm bu}$ vary accordingly. Nevertheless, the proportionality factor
\begin{equation}
f\equiv \sigma_{\rm bu} / b^2
\label{prop}
\end{equation}
does not change because of the peripherality. Therefore, $f$ can be used as a measure of the peripherality.

\begin{figure}
\includegraphics[scale=1.1]{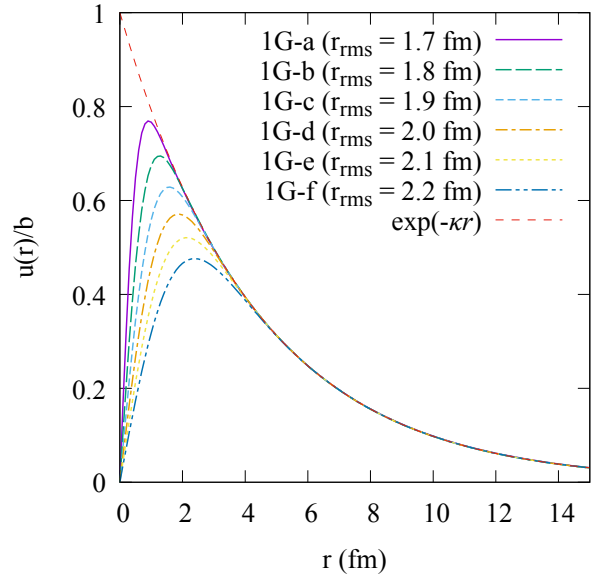}
\caption{\label{fig:tail}
Deuteron wave functions normalized to the exponential function at 6~fm.
}
\end{figure}

\begin{table}[hptb]
\caption{\label{tab:table1}%
One-range Gaussian potentials prepared for the peripherality study. Parameters of 1G-av4 are also listed. $V_0$ is determined to reproduce the binding energy calculated with the AV4' potential.
}
\begin{ruledtabular}
\begin{tabular}{cccc}
\textrm{name}&
\textrm{$V_0$ (MeV)}&
\textrm{$a$ (fm)}&
$r_{\rm rms}$ (fm)\\
\colrule
1G-a & 280.23 & 0.687 & 1.70\\
1G-b & 131.42 & 1.047 & 1.80\\
1G-c & 79.21 & 1.405 & 1.90\\
1G-d & 54.35 & 1.765 & 2.00\\
1G-av4 & 52.10 & 1.812 & 2.01\\
1G-e & 40.46 & 2.126 & 2.10\\
1G-f & 31.76 & 2.492 & 2.20
\end{tabular}
\end{ruledtabular}
\end{table}
We prepare six one-range Gaussian potentials which generate deuteron wave functions with rms radii ranging from 1.7 fm to 2.2 fm. Their depth and range parameters are shown in Table~\ref{tab:table1}. Figure~~\ref{fig:tail} represents the resulting deuteron wave function divided by $b$; $b$ is extracted at 6~fm.

\begin{figure}[ht]
\includegraphics[scale=1.]{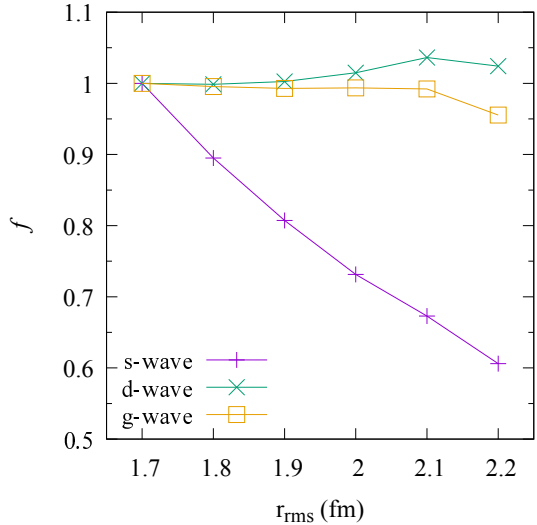}
\caption{$f$ for each partial-wave component of the breakup cross section on $^{58}$Ni at 80~MeV. The horizontal axis is the rms radii of deuteron adopted. The values of $f$ are normalized to the value at $r_{\rm rms}=1.7$~fm.
}
\label{fig:80MeV_b2sigma_full}
\end{figure}

Figure~\ref{fig:80MeV_b2sigma_full} shows $f$, normalized to the value at $r_{\rm rms}=1.7$~fm, for each partial-wave component of the breakup cross section on $^{58}$Ni at 80~MeV. For the d- and g-wave breakup, $f$ is almost constant, which indicates the peripherality of the reaction. On the other hand, $f$ for the s-wave breakup strongly depends on $r_{\rm rms}$. This means that the s-wave deuteron breakup is not peripheral. 
\begin{figure}
\includegraphics[scale=1.]{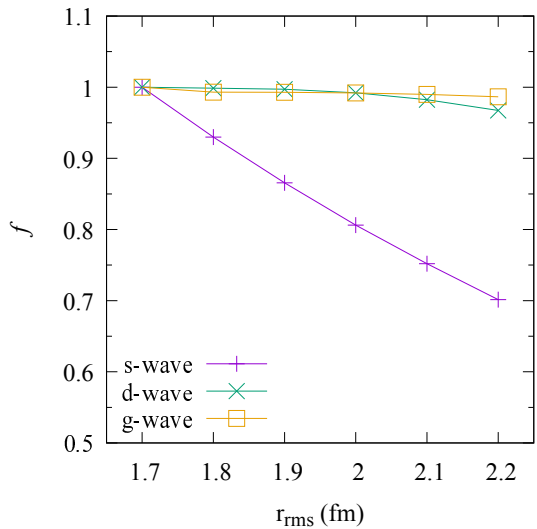}
\caption{\label{fig:b2sigma_1GeV}
Same as Fig.~\ref{fig:80MeV_b2sigma_full} but at 1~GeV.
}
\end{figure}
In Fig.~\ref{fig:b2sigma_1GeV} we show the result at 1~GeV. The general feature is the same as at 80~MeV but the $r_{\rm rms}$ dependence of the s-wave breakup is slightly weaker. It is found that this weakening is due to the less importance of the multistep breakup processes at 1~GeV. In other words, at 80~MeV, multistep processes enhance the contribution from the inner part of deuteron.

Thus, the negligible difference between the breakup cross sections with the AV4' and 1G-av4 interactions shown in Figs.~\ref{fig:80MeV_pbu_micsys} and \ref{fig:1GeV_pbu_micsys} can be understood by the peripherality of the reaction, except for the s-wave breakup. On the other hand, the s-wave breakup is found to be not peripheral, which appears to contradict with the finding shown in Figs.~\ref{fig:80MeV_pbu_micsys} and \ref{fig:1GeV_pbu_micsys}. A possible way of understanding the phenomenon will be that the whole spatial region of deuteron is probed and the difference between the deuteron wave functions with the AV4' and 1G-av4 interactions is smeared. As shown in Fig.~\ref{fig:pot-u_r-2.01}, the solid line is larger than the dashed line between 1~fm and 2~fm, whereas the former is smaller than the latter at $r<1$~fm. This may indicate that if a breakup process that selectively probes $r$ larger than 1~fm was found, it could be a probe of the short-range repulsive core of $V_{pn}$.

\section{Conclusion\label{sec:conclusion}}

We have investigated the effect of the short-range repulsive core of the $p$-$n$ interaction on the deuteron breakup cross sections on $^{58}$Ni at incident energies from 40~MeV to 1~GeV. While the deuteron wave function is affected by the repulsive core at the $p$-$n$ distance $r$ less than 2~fm, the deuteron breakup cross section change very little. This insensitivity is found to be due to the peripherality of the reaction process concerning $r$ except for the s-wave breakup. The s-wave breakup is found to be non-peripheral. The insensitivity of the s-wave breakup cross section to the short-range repulsive core may suggest that this reaction probes the whole spatial region of deuteron. The exact extent and mechanism of the non-peripheral characteristic of the s-wave breakup will need further investigation.

\begin{acknowledgments}
The authors would like to thank Shin Watanabe and Jagjit Singh for fruitful comments and discussions. This work was supported in part by Grants-in-Aid of the Japan Society for the Promotion of Science (Grant No. JP16K05352).
\end{acknowledgments}

\bibliography{ref}

\end{document}